# THE STATISTICAL INVESTIGATION OF TYPE Ib/c AND II SUPERNOVAE AND THEIR HOST GALAXIES

**A. A. Hakobyan**
(V. A. Ambartsumian Byurakan Astrophysical Observatory, Armenia; e-mail: hakartur@rambler.ru)

*This is a statistical study of the properties of type Ib/c and II supernovae and of the integral parameters of their spiral host galaxies. The methods of one-dimensional and multivariate statistics were applied to the data sample. It was found that the Ib/c supernovae are more concentrated radially toward the centers of the galaxies than those of type II. The distributions of the radial distances $R_{SN} / R_{25}$ for the type Ib/c and II supernovae in active galaxies are more concentrated toward the center than in normal galaxies. This effect is stronger for type Ib/c than for type II supernovae.*

## 1. Introduction

Studies of supernovae and their host galaxies provide important keys to our understanding of their chemical evolution, the kinematics of the interstellar medium, the origin of cosmic rays, the history of the formation and subsequent stages of evolution of stars, and the possible nature of the progenitors which evolve into supernova bursts of various kinds. For these purposes, studies of type Ia supernovae [1] and of types Ib/c and II supernovae and their host galaxies [2] have different goals. Type Ib/c and II supernovae and their host galaxies are of special interest and are the subjects of the present study. Many papers have been published about these objects [3,4]. Type II supernovae bursts with the core collapse of young, massive stars - the red supergiants, whose envelopes contain a large amount of hydrogen. Type Ib/c supernovae also bursts during collapse of the cores of massive stars, probably Wolf-Rayet stars [4], but a large portion of the hydrogen envelopes of these stars are lost during the evolution of a close binary system or by some other means, which is probably why the progenitors of Ib/c supernovae have a higher metallicity [3,5]. The research done thus far on the relationships among the properties of these supernovae and their host galaxies can be summarized as follows [3,6].

The properties of supernovae depend on the Hubble morphological type of the galaxies in which they have burst. Type Ib/c and II supernovae have only been observed in spiral galaxies, where stars are still being formed and both old, low mass stars and young, massive stars are present. No significant difference has been found in the distributions of the morphological type of the host galaxies for type Ib/c and II supernovae. This confirms that supernovae in both these classes had similar progenitors [7]. Ib/c and II supernovae are both clearly concentrated in the spiral arms and are related to the disk population of the spiral host galaxies [8]. Supernovae which show a correlation with the spiral arms can have different observational properties from those of supernovae which have no correlation with the spiral arms. The difference originates mainly in the more dense interstellar gas of the spiral arms. The interaction of supernovae with this gas shows up as various effects in their spectra [6]. In addition, Ib/c and II supernovae are closely related to HII regions or to clusters of bright blue stars, the OB associations [9]. The similarities which relates Ib/c and II supernovae to HII regions also implies that the Ib/c and II supernovae had young progenitors within the same range of masses [8,10].

The observed ratio of the number of type Ib/c to type II supernovae depends on the luminosity and metallicity of the host galaxies. The ratio $N_{Ib/c} / N_{II}$ increases with the luminosity of the host galaxy [2].

Differences have been observed between the distributions of supernovae in normal and active host galaxies [11,12]. The radial distribution of type Ib/c supernovae in their host galaxies is more centered than for type II supernovae. The frequencies of type Ib/c and II supernovae depend strongly on the color index ($B$-$K$) of the host galaxy, i.e., on its star formation activity [13].

A relation has been observed between the properties of supernovae and the inclinations of their host galaxies. The rate of supernovae in host galaxies of morphological type Sc-Sd has a peak for objects with inclination of ~25° or less [14].

This paper is a study of the relation between the integral properties of host galaxies and the properties of type Ib/c and type II supernovae observed in them. This can be used to determine the nature of the progenitors of Ib/c and II supernovae. The present study is based on a sample of spiral galaxies with radial velocities ≤ 10000 km s$^{-1}$ and disk inclinations ≤ 50°, and a sample of type Ib/c and II supernovae observed till to August 2006.

Section 2 discusses the database for this study and the methods used to analyze the data. The results and a discussion are given in Section 3. Section 4 is the conclusion. In this paper the Hubble constant is taken to be $H_0$ = 75 km s$^{-1}$ Mpc$^{-1}$.

## 2. Database and data processing

The main source of data was the Asiago electronic catalog of supernovae, which is available at http://web.pd.astro.it/supern/ [15]. To the extent possible, this catalog gives the morphological types, the logarithm $D_{25}$ of the diameters, radial velocities, inclinations of the disks, position angles, magnitudes of the host galaxies in blue light (usually from RC3 or the Leda database), the coordinates of the supernova relative to the center of the host galaxy, the supernova types, etc. The main sources for data on the supernovae in the catalog are the IAU circulars. For the present study, we have only chosen supernovae in subtypes Ib, Ic, Ib pec, and Ic pec as general type Ib/c, and subtypes IIb, IIn, IInL, IIL, IIB-L, IIP, IIB-P, and II pec as general supernova type II. The latest supernova included in this study was SN 2006ep (date of discovery, August 30, 2006). The chosen supernovae and their host galaxies were those for which the following data are listed in the catalog: (1) the coordinates of the supernova relative to the center of host galaxy, (2) the morphological type of the host galaxy (spiral), (3) the radial velocity of the host galaxy (≤ 10000 km s$^{-1}$), (4) the inclination of the host galaxy, (5) position angle of the host galaxy, (6) the blue magnitude of the host galaxy, and (7) the logarithm of the diameter $D_{25}$ ($D_{25} = 2R_{25}$). In order to minimize absorption and projection effects, host galaxies with $i > 50°$ were excluded. As a result, only 271 supernovae and 243 host galaxies (68 type Ib/c and 203 type II supernovae) out of the 3618 supernovae observed up to August 2006 satisfy these conditions. In each of the 221 host galaxies, at least one of these supernovae bursts: in 17 host galaxies two supernovae burst, in 4 host galaxies three burst, and in one host galaxy 4 burst. Here the radius $R_{25}$ is taken to be the major semiaxis of the host galaxy to a distance with a limiting brightness of 25 magnitudes from one arc second squared. It was assumed that all the Ib/c and II supernovae burst on the plane of the disks of the host galaxies and the radial distances $R_{SN}$ from the centers of their host galaxies were calculated taking the inclination of the disk plane into account.

The source of the data on the infrared $K$ (2.17 m) magnitudes of the host galaxies was the electronic archive of the 2 micron all sky survey (2MASS), which is accessible at http://irsa.ipac.caltech.edu/. This archive gives $K$ magnitudes for only 208 of the 243 host galaxies. The source of the data for the magnitudes of 192 of the host galaxies (out of the 243) at a wavelength of 21 cm (1.42 GHz) was the HyperLeda database of extragalactic objects, which is accessible at http://leda.univ-lyon1.fr/. Finally, all of these data were accessible for only 175 supernovae and 161 host galaxies out of the 271 supernovae and 243 host galaxies.

Data from the NASA/IPAC (NED) data base, which is accessible at http://nedwww.ipac.caltech.edu/, were used to determine the activity class of the host galaxies. The host galaxies in the sample were assumed to be active if (1) their nuclei indicated different levels of activity (Sy1, Sy1.5, Sy2, LINER, LINER/HII, etc.) and (2) if star-formation (starbursts, HII) is evident. Of the sample of host galaxies, 42 (17%) were active and 201 (83%) were normal. In the active host galaxies there were a total of 52 (19%) supernovae (19 (7%) Ib/c and 33 (12%) II) and in the normal host galaxies a total of 219 (81%) supernovae (49 (18%) Ib/c and 170 (63%) II).

The data were analyzed statistically using the methods of one-dimensional and multivariate statistics, in particular, Multivariate Factor Analysis (MFA). The MFA statistical method is used to discover correlations among a series of $m$ initial variables measured for $n$ objects and the linearly independent factors $F_1, F_2, ..., F_k$ ($k < m$). The main goals of MFA are (1) to reduce the number of variables and (2) to determine the structure of the relation among the variables, i.e., classify or group the variables. This method has been used in astronomy [16-19]. A detailed description of the MFA method is given elsewhere [20,21].

When using the MFA it is appropriate to use two samples of the initial variables. In the first sample, for the 271 supernovae and their 243 host galaxies, the following initial variables were used: morphological type (Morph: S0 = -2, S0/a = -1, S = 0, Sa = 1, Sab = 2, Sb = 3, Sbc = 4, Sc = 5, Scd = 6, Sd = 7, Sdm = 8), bar parameter ($B$: 0 when no bar is present in the host galaxy and 1 when bars are present), disk inclination ($i$), absolute blue magnitude of the host galaxy ($M_B$), radius $R_{25}$ (kpc), supernova type parameter ($T$: 0 for type Ib/c and 1 for type II), relative radial distance $R_{SN} / R_{25}$ from the center of the host galaxy taking the disk inclination into account, and activity parameter (A/SF: 0 for normal and 1 for active galaxies). In the second sample, for 175 supernovae and their 161 host galaxies, the above variables were supplemented by the absolute magnitude at 21 cm ($M_{21}$) and the color index ($B$-$K$) of the host galaxies.

### 3. Results and discussion

In the sample of all 271 supernovae, the ratio of type Ib/c to type II supernovae ($N_{Ib/c} / N_{II}$) is ~0.33. A theoretical model [22] indicates that $N_{Ib/c} / N_{II}$ never exceeds ~0.2, which is not confirmed with observations [2, 23]. It has been found that in normal galaxies $N_{Ib/c} / N_{II} \sim 0.27$ [23], while it has been claimed [24] that one out of four type Ib/c and II supernovae belongs to type Ib/c, so that $N_{Ib/c} / N_{II} \sim 0.33$. These data are consistent with the result in this paper.

Table 1 lists the correlation coefficients between the variables and the different factors for the 271 supernovae and their 243 host galaxies. In order to simplify the interpretation of the results, a special method was used for rotating the factors. This type of rotation is known as rotation that maximizes the dispersion (varimax), since the purpose of the rotation is to maximize the dispersion (variability) of a variable (factor) and minimize the scatter around it. In this case, taking a correlation threshold of ~0.4 for the first factor $F_1$, which explains ~21.6% of the total dispersion, we find that it groups the absolute magnitude $M_B$, the radius $R_{25}$ (kpc), and morphological type of the host galaxies. These dependences do not concern the properties of the supernovae, but it is known from the literature that galaxies with higher luminosities have larger linear sizes and that the luminosity decreases systematically for galaxies over the morphological types from S0 to Sdm [25]. The factor $F_2$, which explains ~16.6% of the overall dispersion, groups the relative radial distance $R_{SN} / R_{25}$, the supernova type, and the activity parameter of the host galaxy. These dependences show that type II supernovae lie farther from the center of the host galaxy than those of type Ib/c. In addition, supernovae of both types lie farther from the center of normal host galaxies than the supernovae in active host galaxies.

TABLE 1. Correlation Coefficients Between the Variables and the Different Factors for 271 Supernovae and their 243 Host Galaxies

| Variable | $F_1$ | $F_2$ | $F_3$ |
|---|---|---|---|
| Morph | 0.414 | 0.042 | 0.396 |
| $B$ | -0.146 | -0.025 | 0.554 |
| $i$ | 0.093 | 0.102 | 0.686 |
| $M_B$ | 0.899 | 0.092 | -0.047 |
| $R_{25}$ (kpc) | -0.833 | 0.066 | 0.082 |
| $T$ | -0.069 | 0.694 | -0.002 |
| $R_{SN} / R_{25}$ | 0.069 | 0.705 | 0.252 |
| A/SF | -0.108 | -0.571 | 0.438 |
| Percent of overall disp. (%) | 21.6 | 16.6 | 15.0 |

To clarify the relation of $R_{SN} / R_{25}$ to the type Ib/c and II supernovae, Fig. 1 shows histograms of the relative radial distances $R_{SN} / R_{25}$ from the centers of the host galaxies. The mean relative radial distances for Ib/c and II supernovae are equal, respectively, to 0.39 ± 0.04 ($N_{Ib/c}$ = 68) and 0.54 ± 0.02 ($N_{II}$ = 203). This means that type Ib/c supernovae are more concentrated toward the centers of the host galaxies than are the type II supernovae (confidence level 98.9%). 51 of the 68 type Ib/c supernovae (75%) lie within $R_{SN} / R_{25}$ = 0.5, by comparison with 103 of the 203 type II supernovae (51%). The higher concentration of type Ib/c supernovae toward the galactic center can be explained by the systematically higher metallicity of the central regions of spiral galaxies [2,26]. For example, it has been found [2] that the ratio $N_{Ib/c} / N_{II}$ increases with increasing metallicity; that is, the fraction of type Ib/c supernovae increases toward the center of the host galaxy.

To clarify the relation of $R_{SN} / R_{25}$ to the activity parameter of the host galaxies, Fig. 2 shows histograms of the relative radial distances $R_{SN} / R_{25}$ from the centers observed in active and normal host galaxies. The averages of $R_{SN} / R_{25}$ for active and normal host galaxies are equal, respectively, to 0.42 ± 0.04 ($N_{SN}$ = 52) and 0.52 ± 0.02 ($N_{SN}$ = 219). This ratio for type Ib/c supernovae in active and normal host galaxies is, respectively, 0.29 ± 0.07 ($N_{Ib/c}^{a} = 19$) and 0.43 ± 0.04 ($N_{Ib/c}^{n} = 49$), while for type II supernovae, the corresponding values are 0.49 ± 0.05 ($N_{II}^{a} = 33$) and 0.55 ± 0.02 ($N_{II}^{n} = 170$).

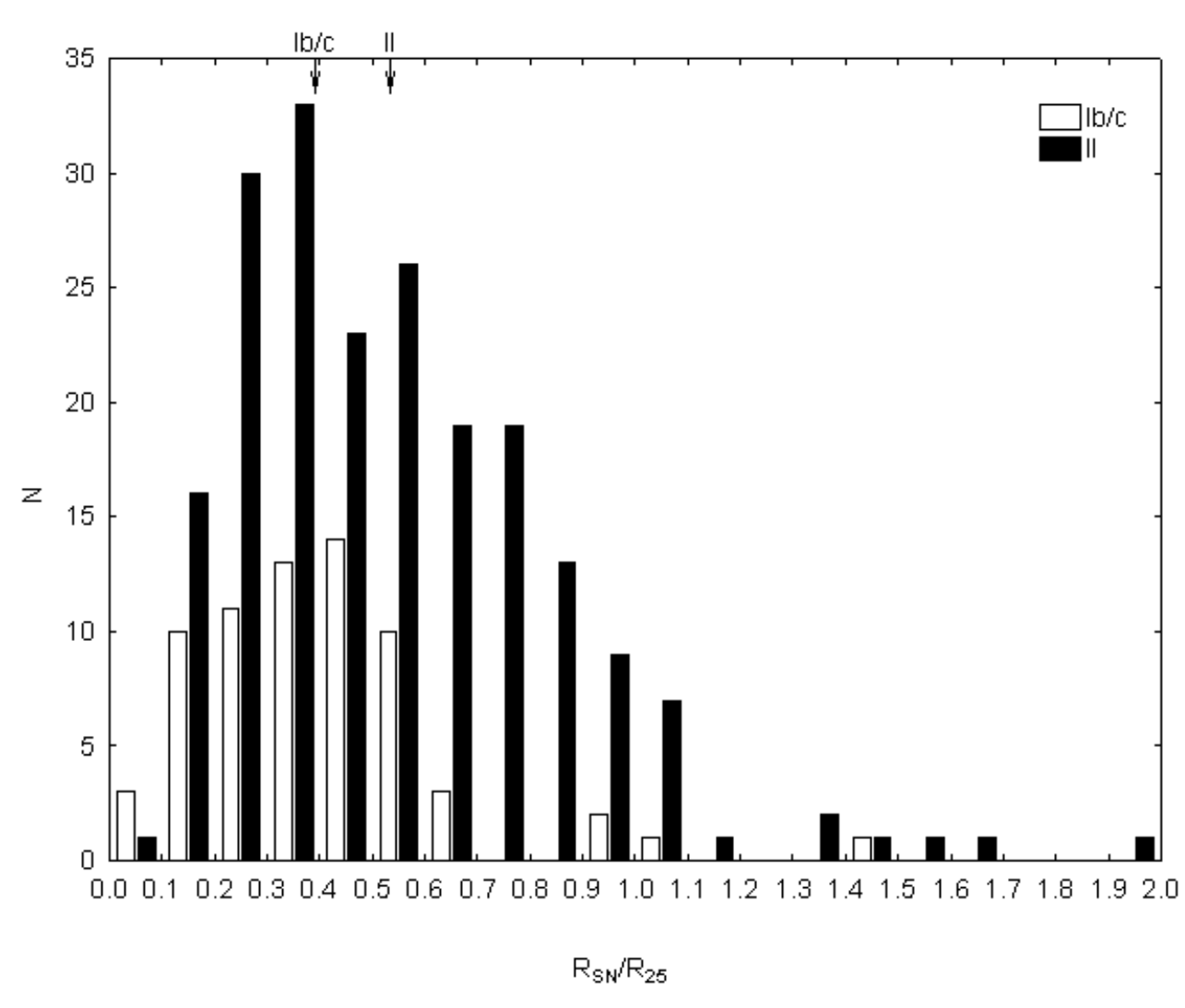

Fig. 1. Histograms of the relative radial distances $R_{SN} / R_{25}$ from the centers of host galaxies for type Ib/c and type II supernovae. The supernovae of classes Ib/c and II are indicated, respectively, in white and black. The average values of $R_{SN} / R_{25}$ are indicated by arrows.

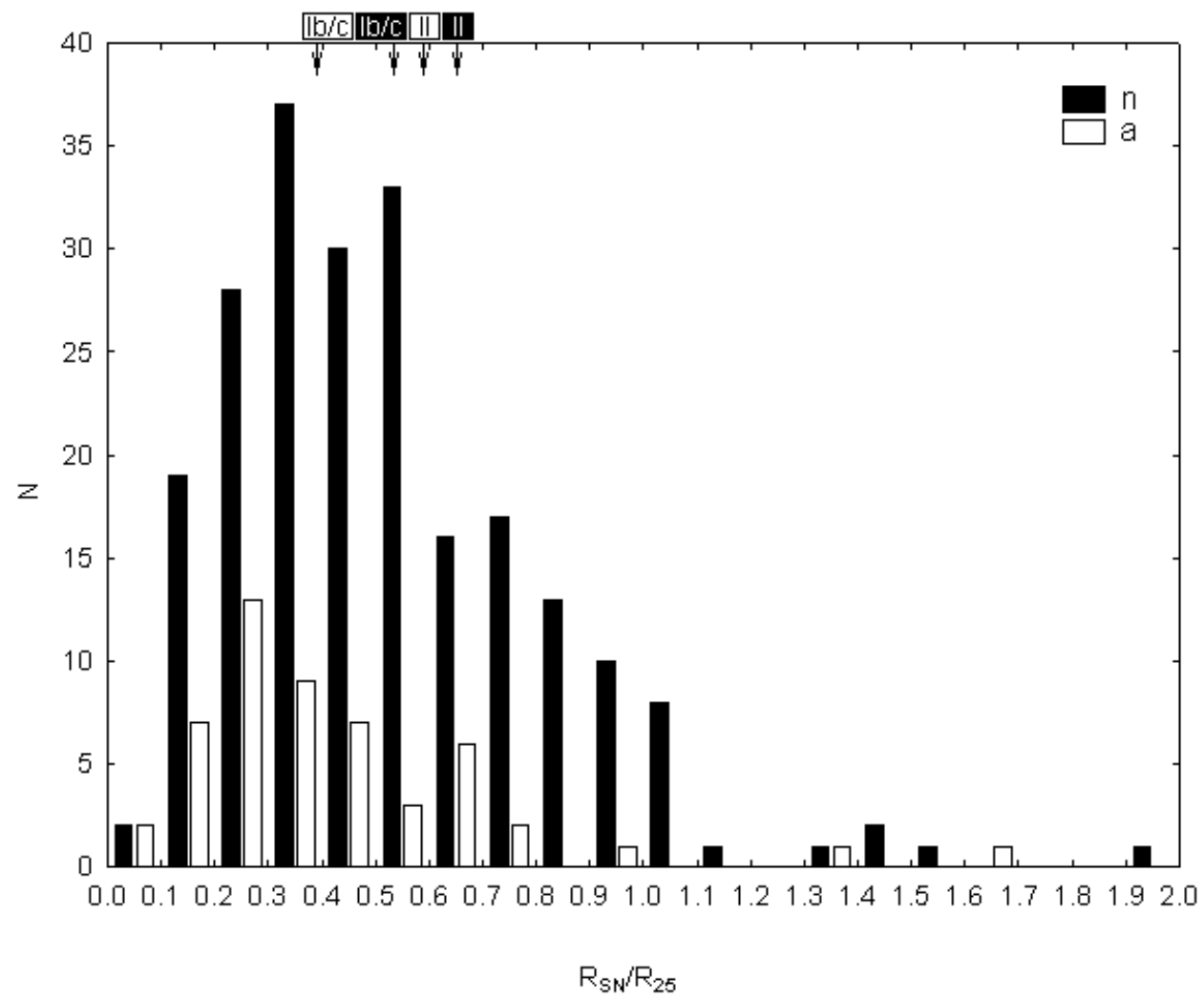

Fig. 2. Histograms of the relative radial distances $R_{SN} / R_{25}$ from the centers observed in active and normal host galaxies. The supernovae in active and normal host galaxies are indicated, respectively, in white and black. The average values of $R_{SN} / R_{25}$ are indicated by arrows.

The higher concentration of supernovae toward the center of active host galaxies (confidence level 91.3%) can be explained by the higher metallicity of the central regions of active galaxies compared to normal galaxies [27]; here this effect is stronger for type Ib/c than for type II supernovae. For example, it has been found [23] that in Seyfert galaxies the ratio $N_{Ib/c} / N_{II} \sim 1$, i.e., the fraction of Ib/c supernovae increases toward the center of active galaxies more rapidly than in the case of normal galaxies.

The factor $F_3$, which explains ~15% of the overall dispersion, groups the bar parameter, inclination of the disk, activity parameter, and morphological type of the host galaxies. It is well known that many of the physical properties of galaxies vary systematically along the morphological sequence of types from S0 to Sdm. In spiral galaxies, activity of the nucleus and/or star-formation is mainly observed in giant galaxies with a bar in later morphological types [25]. In addition, on the average, galaxies in the later morphological types have higher disk inclinations [28].

Table 2 lists the correlation coefficients between the variables and the different factors for the 175 supernovae and their 161 host galaxies in the second sample. The first factor $F_1$, which explains ~23.1% of the overall dispersion, groups the absolute magnitude $M_B$, the radius $R_{25}$ (kpc), and the absolute magnitude of the host galaxy at a wavelength of 21 cm. Here another well known result stands out: in giant galaxies with a high luminosity, the abundance of neutral hydrogen radiating at 21 cm is systematically higher [25]. The factor $F_2$, which explains ~14.0% of the overall dispersion, groups the morphological type, disk inclination, and color index ($B$-$K$) of the host galaxies. It is well understood that on going from galaxies of morphological type S0 to type Sdm, the color of the galaxies becomes bluer and, as noted above, the inclination of their disks increases [28]. The factor $F_3$, which explains ~13.2% of the overall dispersion, groups the variables in roughly the same way as factor $F_2$ in the first sample.

TABLE 2. Correlation Coefficients Between the Variables and the Different Factors for 175 Supernovae and their 161 Host Galaxies

| Variable | $F_1$ | $F_2$ | $F_3$ |
|---|---|---|---|
| Morph | 0.003 | 0.727 | -0.151 |
| $B$ | 0.191 | 0.081 | 0.120 |
| $i$ | 0.092 | 0.512 | 0.228 |
| $M_B$ | 0.818 | -0.003 | 0.013 |
| $R_{25}$ (kpc) | -0.890 | -0.086 | 0.055 |
| $T$ | 0.040 | -0.146 | 0.653 |
| $R_{SN} / R_{25}$ | 0.102 | 0.238 | 0.756 |
| A/SF | 0.140 | 0.096 | -0.443 |
| $M_{21}$ | 0.878 | 0.021 | -0.142 |
| $B$-$K$ | -0.067 | -0.714 | 0.121 |
| Percent of overall disp (%) | 23.1 | 14.0 | 13.2 |

## 3. Conclusion

The relation between the properties of supernovae of types Ib/c and type II and the integral parameters of their host galaxies have been examined in this paper. The major results are the following:

- Type Ib/c supernovae are more concentrated toward the center of the galaxy than are type II supernovae. This is consistent with the high metallicity of the central regions of the galaxies.
- The radial distributions of type Ib/c and type II supernovae in active galaxies are more concentrated toward the center than in normal galaxies. This effect is stronger for type Ib/c than for type II supernovae, and is an indicator of a higher rate of star formation in the regions near the nuclei of active galaxies.

These data are consistent with generally accepted ideas that the progenitors of supernovae of types Ib/c and II are associated with the young star population of galaxies [23,29].

This paper has used data from the Asiago supernova catalog, the Lyon-Medon database of extragalactic objects (HyperLeda) supported by the LEDA group and the CRAL observatory in Lyon, and the NASA/IPAC infrared data archive and the NASA/IPAC database of extragalactic objects (NED) operated by the Jet Propulsion Laboratory of Caltech under NASA contract.

The author thanks A. R. Petrosian for his support and critical comments, which have substantially improved this article. This work was supported by a grant from the French government.